\input{aipcheck}

\documentclass[
    ,final            
  ]
  {aipproc}

\layoutstyle{8x11double}

\begin{abstract}

A method to search for particles of unknown masses  in events with
missing energy is presented. The only assumption is the topology of the decay chain.
The solvability of the event has  significant mass information and can be used to
extract the unknown masses. The method is tested in events with top pairs decaying
leptonically with the presence of two neutrinos in the final state.
Similar topologies are predicted by supersymmetric models  with conserved R-parity
were sparticles are produced in pairs through  cascade  decay chains with the
presence of the lightest neutralino in the end of each chain.

\end{abstract}

\begin{document}

\title{Model Independent Search in 2-Dimensional Mass Space }


\author{G.~Anagnostou }{
  address={I. Physikalisches Institut B, RWTH Aachen, Germany }
}

\begin{abstract}

A model independent method to search  for particles of unknown masses  in events with
missing energy is presented. 
The only assumption is the topology of the decay chain.
The method is tested in events with top pairs decaying leptonically with the presence 
of two neutrinos in the final state. Possible applications are searches for a heavy
resonance decaying to top pairs as well as next generation heavy quarks.
Similar topologies are predicted by supersymmetric models  with  R-parity conservation
resulting in final states with two invisible neutralinos.
 
\end{abstract}

\maketitle



\begin{figure}[t!]
\includegraphics[scale=0.385]{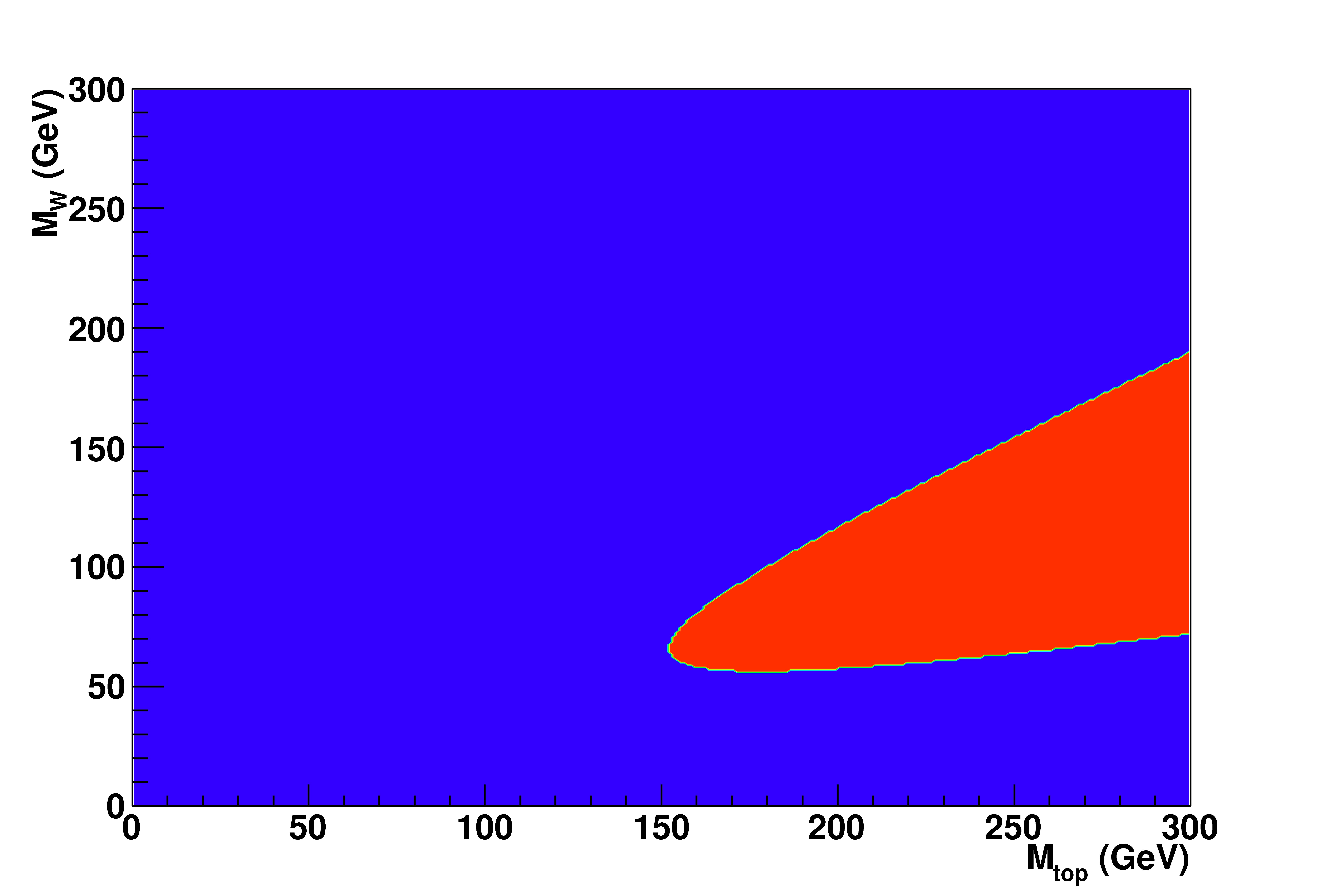}
\includegraphics[scale=0.385]{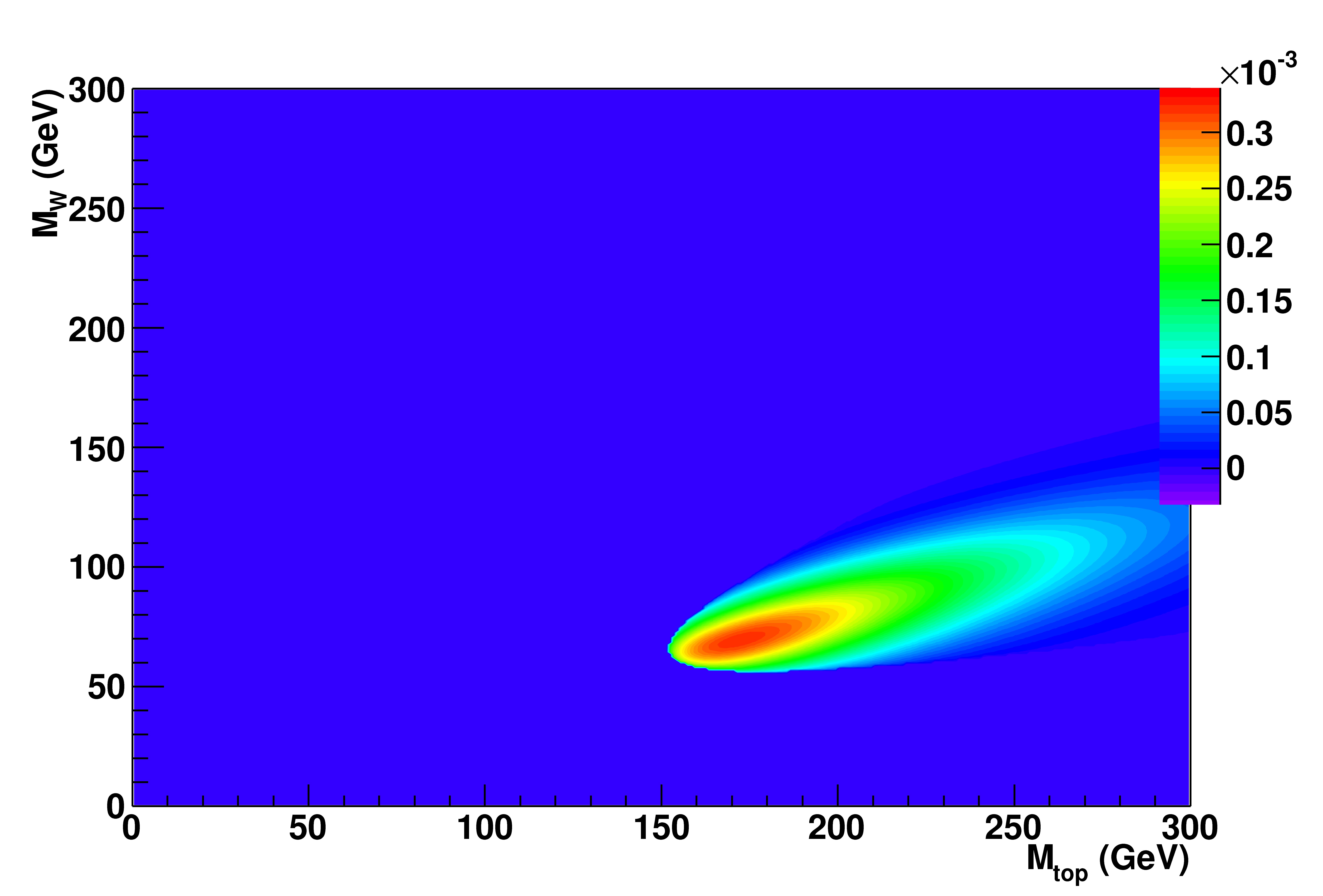}
\caption{
(Left): Solvability for the correct solution of  a single top-pair event in the 2-dimensional mass plane.
(Right): Solvability weighted with the PDFs and normalized to unit volume for the same solution and event. 
}
\label{fig:singleevent_sol}
\vspace{2em}
\end{figure}

\begin{figure}[t!]
\includegraphics[scale=0.385]{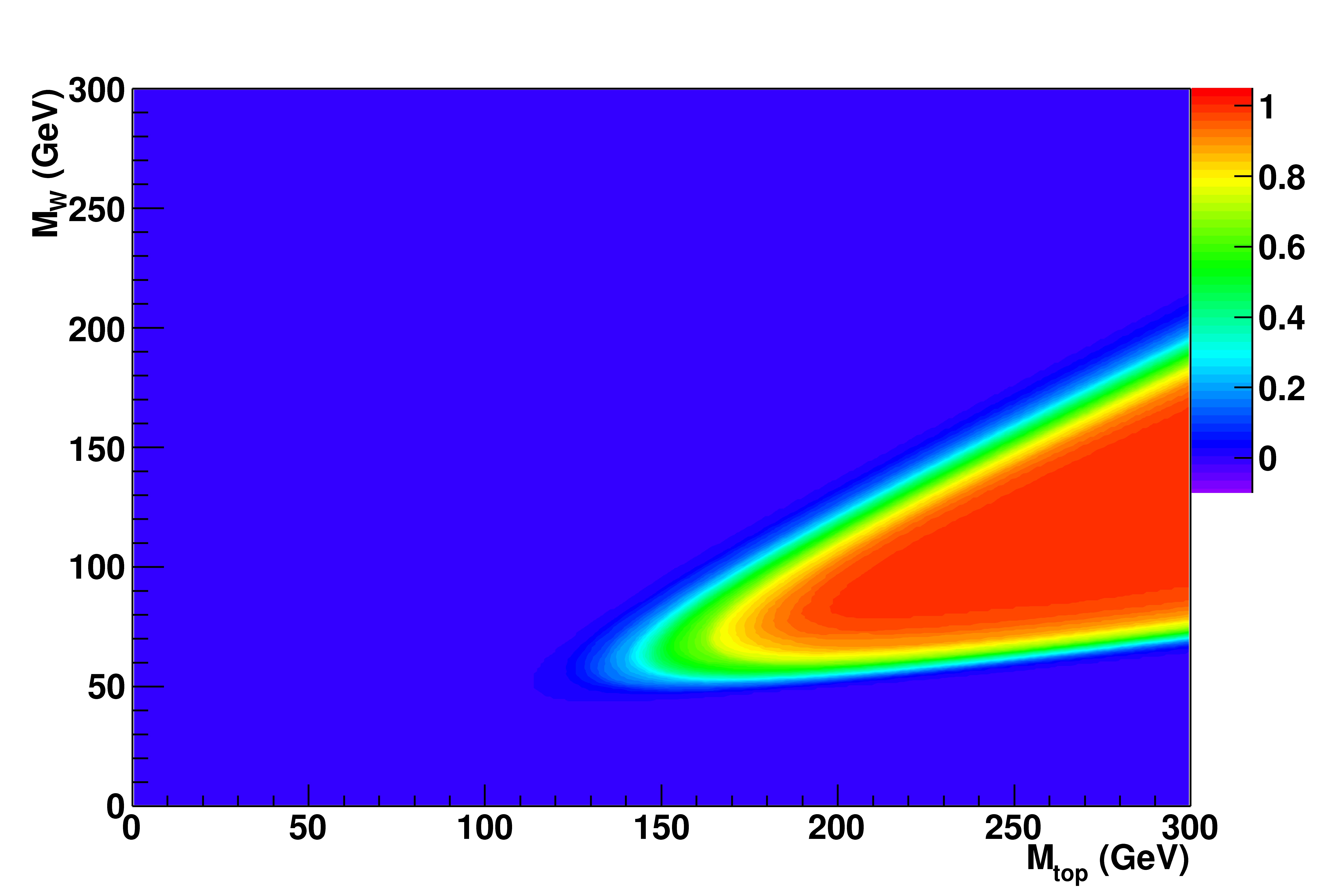}
\includegraphics[scale=0.385]{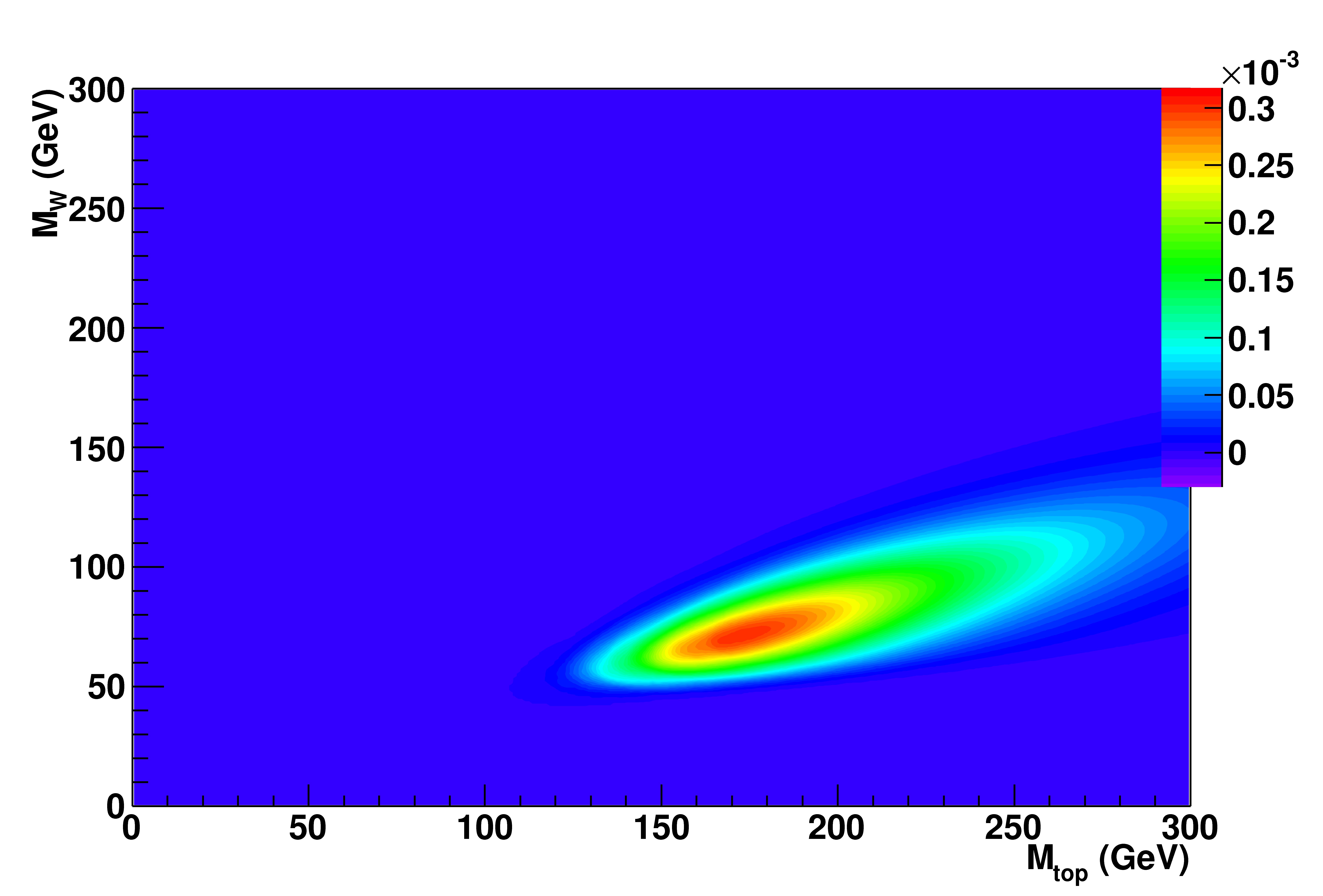}
\caption{
(Left): Solvability (for the correct solution) of 300 test events produced from the same single event by smearing of the momenta of leptons, jets and
transverse missing energy.
(Right): Average weighted solvability for the same test events normalized to unit volume.
}
\label{fig:singleevent_300}
\vspace{2em}
\end{figure}

\section{Introduction}

 Could we find the mass of the top quark and W boson in LHC from top-pairs decaying
leptonically in the hypothetical case in which both $\mathrm{m_{t}}$ and $\mathrm{m_{W}}$  were unknown?
Could we establish a discovery by observing the mass peaks above background for both 
particles without any assumption about the underlying theory?

Standard Model has predicted the mass of the W boson which was later measured
accurately by LEP. Top mass has also  been measured in both Tevatron and LHC.
The reason that such a question is interesting is the similarity with BSM  theory models
which predict new invisible particles (e.g neutralinos). It is believed by many physicists
that such theories might be the
next step in our understanding of the physical world  as they provide (among others) an explanation for the nature
of dark matter. For example, supersymmetric models with R-Parity conservation
predict  event topologies with at least two neutralinos.
The existance of two invisible particles in the decay chain makes the extraction of the unpredicted by
the model masses a difficult problem \cite{m1}, \cite{m2}, \cite{m3}. 
Similar topologies might exist in the decays of a hypothetical heavier quark from a fourth generation or
a heavy resonance decaying to top pairs.

Observation of an excess of events with large missing energy  would
provide evidence for the existance of new physics but could not help
further in our understanding of what the new physics is.
 Several $\mathrm{M_{ET}}$-like observables have been proposed for this purpose 
($\mathrm{M_{ET}}$, $\mathrm{H_{T}}$, $\mathrm{M_{eff}}$ etc).
A common characteristic of all of them is that new physics would appear 
as an excess of events in the tail of a distribution. 
The establishment of a discovery would be a difficult problem  due to the small difference
in the shape between signal and background. In this case, 
a very good understanding of the tail would be necessary,  a challenging task at least at
an early stage.
But let's say that LHC experiments observe an excess of events in the tail of
a $\mathrm{M_{ET}}$-like observable and have confidence in the result. What can
we say about new physics except that it exist?

Understanding the physics is much easier in possible BSM signals which do not predict any invisible
particles. In this case, the invariant  masses can be reconstructed  up to  combinatoric ambiguities.
Knowledge of the masses provides full understanding of the event kinematics 
and subsequently allows  boosts to the correct rest frames of the decaying particles.
Angular distributions in these rest frames  provide the best observables to discriminate between
different spin hypothesis. Knowledge of both mass and spin of the new particles is a very
important step towards understanding what the new physics is.

In addition, signal events are contained in a small region of  mass space in contrast to the higher 
in dimensions kinematic phase space in which selection cuts are often applied.
For example, in the supersymmetric pseudo-scalar MSSM Higgs decay
$\mathrm{A \rightarrow Zh, Z \rightarrow l^{+}l^{-}, h \rightarrow b\bar{b}}$ the reconstruction of both A,h hypothetical Higgs bosons is possible.
The new physics events would then be observed as a peak in the 2-dimensional
$\mathrm{ m_{A}, m_{h}}$ mass plane \cite{anagnostou}.
In this case, there is significant difference in shape between signal and background. 
The establishment of a discovery is an easier task as a data-driven estimation of 
the background can be performed from the sidebands.

Searching for resonances in mass-space is model independent as the signal
extraction does not need any change that depends on the hypothetical model.
The selection applied can be simple energy thresholds to ensure correct reconstruction as 
well as  identification of the  final state objects.

It is often believed that reconstruction of  particles masses in
topologies with two invisible particles is not feasible. In the rest
of this study a counter example is presented with the simultaneous
reconstruction of W boson and top quark in the dilepton top-pair final
state in a p-p collider like LHC. The method can also be used to search
in a model independent way for a new generation $\mathrm {t'}$ and any resonance decaying
to top pairs $\mathrm{pp\rightarrow X'\rightarrow t\bar{t}}$.

\section{Event Simulation and Selection}

All Monte Carlo sample used in this study were generated using Pythia 6 \cite{pythia6}
except the production of a next generation quark via $\mathrm{pp\rightarrow f\bar{f}, q\bar{q} \rightarrow t'\bar{t'}}$ 
for which Pythia 8 was used \cite{pythia8}.
Top pair events with  $\mathrm{m_{t}}$ = 172.5 GeV and  $\mathrm{m_{W}}$ = 80.4 GeV 
were generated corresponding to an integrated luminosity of 1 $\mathrm{fb^{-1}}$. 
For the same intergrated luminosity samples of Drell-Yan   ($\mathrm{Z\rightarrow e^{+}e^{-},\mu^{+}\mu^{-}}$) 
and diboson  backgrounds ($\mathrm{WW}$, $\mathrm{WZ}$, $\mathrm{ZZ}$) were also created.

The performance of the method to new resonances decaying to top pairs
was tested using a  sample of 300 $\mathrm{pp\rightarrow Z'\rightarrow t\bar{t}}$ events,
with  the new gauge boson generated with $\mathrm{m_{Z}}$ = 500 GeV and $\mathrm{\Gamma_{Z}}$ = 15 GeV.
Finally, for the search for a new generation quark, a sample of 300 
$\mathrm{pp\rightarrow f\bar{f}, q\bar{q} \rightarrow t'\bar{t'}}$
events was produced with  $\mathrm{m_{t}}$ = 250 GeV.
Pythia cross-sections were used in all cases except top pairs for which  a more accurate calculation  
was used  \cite{kidonakis}.

\subsubsection {Jet reconstruction}
All stable particles except neutrinos were used as an input to Fastjet-2.4.2 in order
to cluster particles into generated jets \cite{fastjet}.
Jet reconstruction was performed with $\mathrm{anti-K_{T}}$ algorithm using a cone of 0.5.
The generated jets were then smeared according to  the jet reconstruction performance of a typical 
LHC detector \cite{cmstdr}.
The ratio  of the reconstructed to the  partonic  jet energy was used to correct the jet energy
scale. This was performed in 3 $\mathrm{P_{T}}$ bins ($\mathrm{E_{T}}$ < 50 GeV, 50 GeV$\le\mathrm{E_{T}}<$150 GeV, 
$\mathrm{E_{T}}\ge$ 150 GeV) and 2 pseudorapidity bins ($\mathrm{|\eta|}<1.5$, 1.5 < $\mathrm{|\eta|}$ < 2.5).
The correction factors were calculated from the top-pair sample using the bjets matched with a bquark.
All jets matched with a bquark were given 50\% chance to be tagged as bjets.

\begin{figure}[t!]
\includegraphics[scale=0.385]{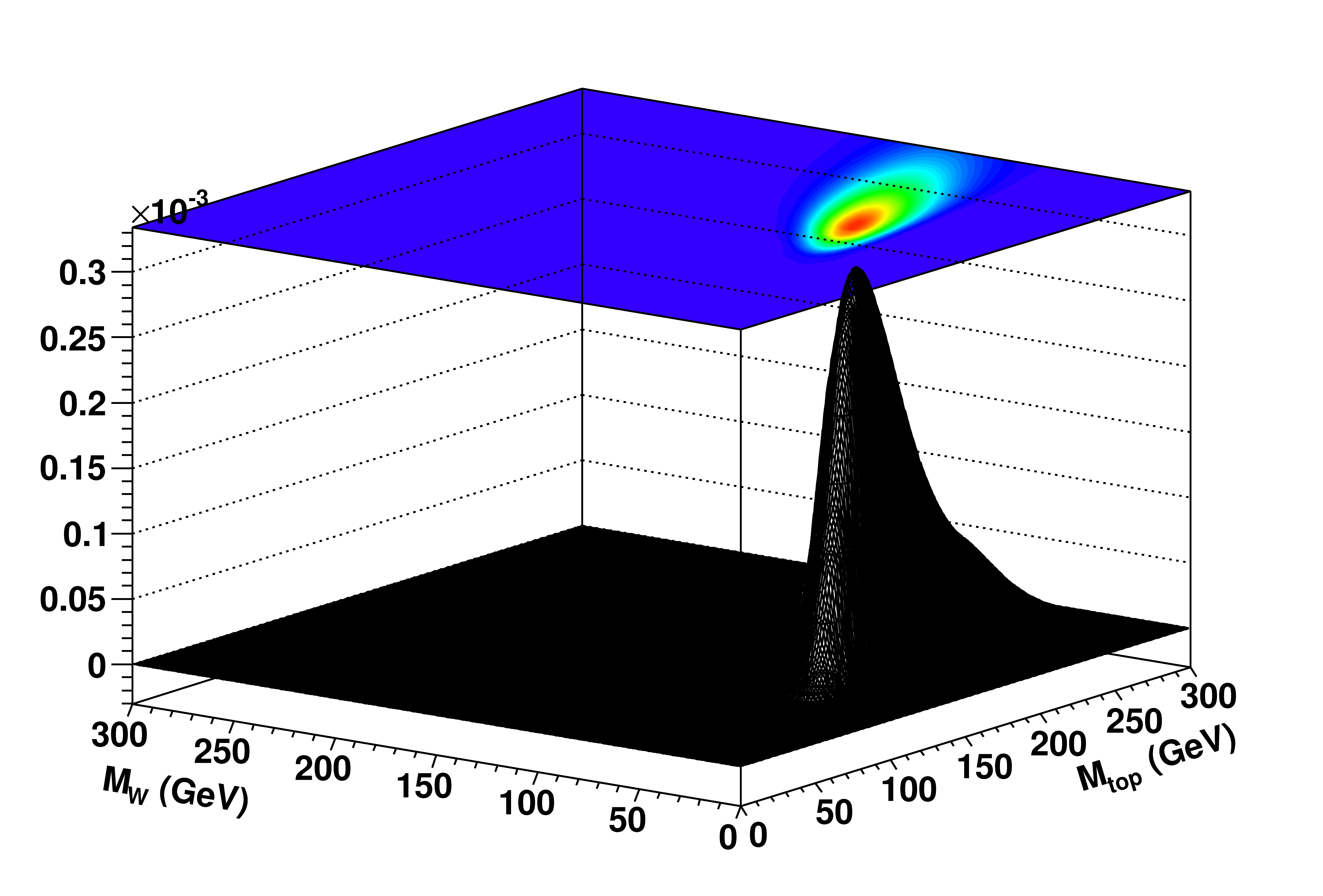}
\caption{
Average  PDF weighted Solvability for the 300  test events produced from the same initial top pair event.
The distribution is  normalized to unit volume.
}
\label{fig:singleevent_3D_300}
\vspace{2em}
\end{figure}

\subsubsection {Leptons and Missing Energy}
Missing energy was calculated by summing vectorially all reconstructed jets,
muons and electrons and then reversing the sign. Both jets and $\mathrm{ M_{ET}}$ resolutions
were checked to be as described in \cite{cmstdr}.
Muon and electron energies were smeared at the percent level which is a typical
resolution for an LHC detector.

\subsubsection {Event Selection}
Events were selected as top pair candidates by requring all objects of the final state
to be above an energetic threshold in order to be well reconstructed:
at least two leptons (electrons or muons) with $\mathrm{P_{T}}\ge$  20 GeV, 
and at least two bjets with $\mathrm{P_{T}}\ge$  20 GeV. In case 
the event had more than two bjets or leptons, the two most energetic ones
were selected.
All jets and leptons were required to be in the pseudorapidity region $\mathrm{|\eta|}$ < 2.5.
 Finally, events with transverse missing energy less than 20 GeV were rejected.
Events satisfying the above selection were used as input to
the algorithm described in the next section.

\begin{figure}[t!]
\includegraphics[scale=0.48]{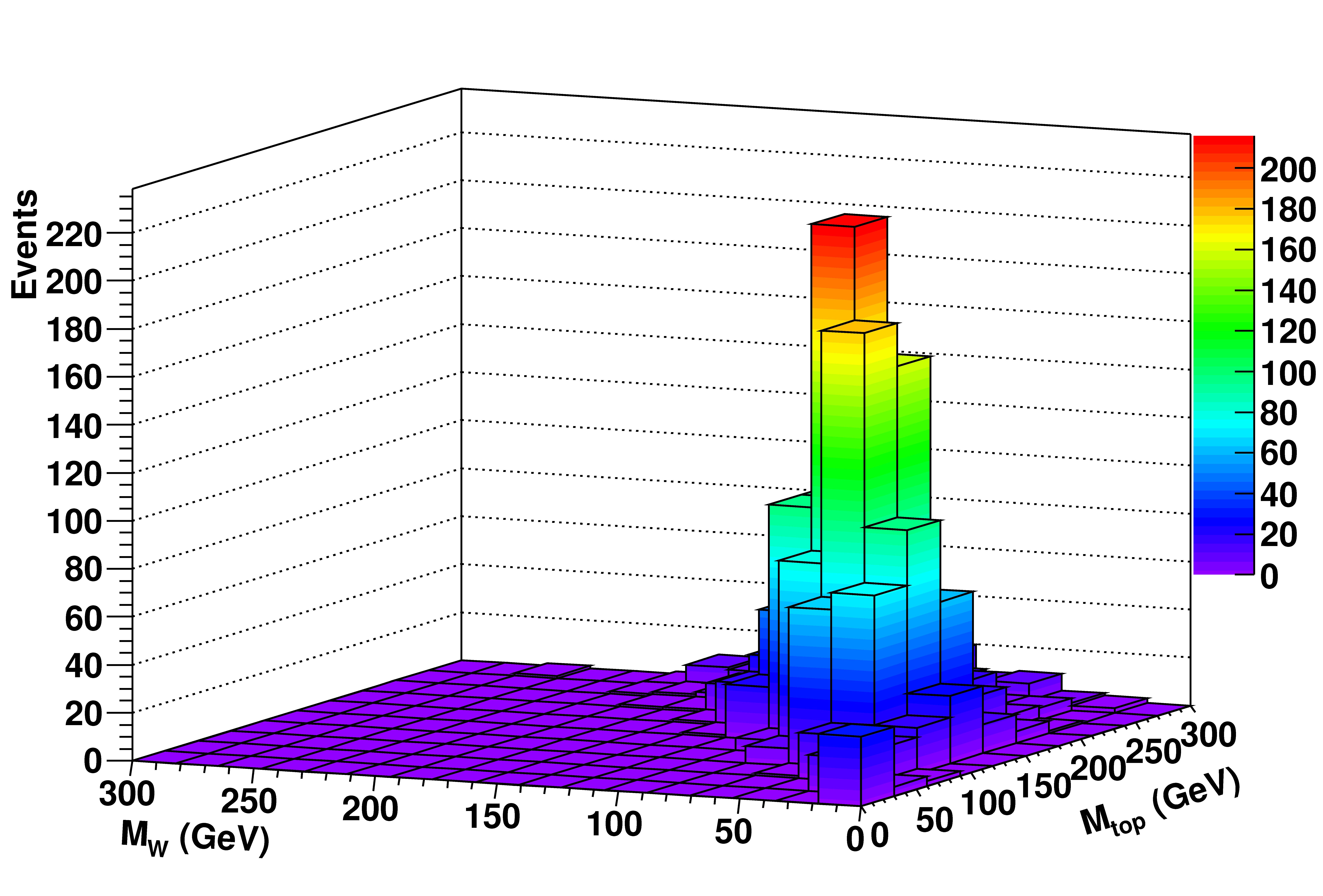}
\caption{
Reconstructed $\mathrm{m_{t}}$ and  $\mathrm{m_{W}}$  per event for a sample of signal and background events
corresponding to an intergrated luminosity of  1 $\mathrm{fb^{-1}}$.
 }        
\label{fig:2D_toppairs}
\vspace{2em}
\end{figure}

\section{The Method}


The kinematics of $\mathrm{ t\bar{t}}$ dilepton events can be expressed by two linear and six non linear equations (Appendix).
The system is solvable with respect to the unknown neutrino and antineutrino four vectors, provided that the masses of 
top quark and W boson are known.
An analytical solution for the equations of the top-pair system is described in \cite{lans1}, \cite{lans2}. 
The transverse missing energy components,  the momentum of bquarks and leptons
as well as the masses of top quark and W boson are used as an input for the analytical solution algorithm.
Each possible input can give 0, 2 or four different solutions for the  unkown neutrino and antineutrino momentum components.
In addition, there are two possible combinations of bjets and leptons that could originate from the same top quark, giving
in total up to eight solutions. So, given the momenta of leptons and bjets together with  $\mathrm{M_{ET_{x}}, M_{ET_{y}}}$
there can be up to  eight possible neutrino and antineutrino momentum vectors for a given $\mathrm{m_{t}}$ and  $\mathrm{m_{W}}$. 
Knowledge of the momenta of the invisible neutrinos allows full kinematic
reconstruction of the event including the four-vectors of W-bosons, top quarks and the $\mathrm{t\bar{t}}$ system.

The initial goal was to "rediscover" W boson and top quark by looking at LHC data without any assumption about
the underlying theory, except the topology of top-pairs. As the masses of the particles are unknown 
the only option is to test  every point of the   $\mathrm{m_{t}}$, $\mathrm{m_{W}}$  plane for possible solutions.
The  mass plane  can be scanned  in steps of 1 GeV to produce the area in which each one
of the solutions exists or not.
Solvability for a single event can be defined as  the existance (or not) of a specific solution  in a specific mass point.
An example of such a solution area for a single top pair event is plotted 
in Figure \ref{fig:singleevent_sol} (left).
By observing several signal events, it can be seen that solvability is  bounded only from below  for both
$\mathrm{m_{t}}$ and $\mathrm{m_{W}}$.

Due to the finite collission energy there is also an upper limit to the allowed masses produced.
The center of mass energy of the partons partipipating in the hard scattering has to be smaller than
the LHC collision energy.
For a p-p collider this energy limit can be expressed fully by the parton distribution functions (PDFs) of the 
proton in the following way: each solution allows full reconstruction of event kinematics, including the 
estimation of the energy E and $\mathrm{P_{Z}}$  momenta component of the $\mathrm{t\bar{t}}$ system. These variables can be easily transformed to
the fraction of beam energy of the two partons participating in the hard scattering ($\mathrm{ x_{1,2}=E\pm pz/2}$).
So each parton with fraction  $\mathrm{x_{i}}$  can be assigned with a probability $\mathrm {F(x_{i}}$) to originate from a proton-proton 
collision. By multiplying the probabilities of the two partons a weight per mass point can be assigned for each solution. 
As there are more than one possible leading order parton-parton interactions
($\mathrm{u\bar{u}}$,  $\mathrm{\bar{u}u}$, $\mathrm{d\bar{d}}$, $\mathrm{\bar{d}d}$,  $\mathrm{gg}$) 
weights from all  possible combinations are summed to estimate a final event weight per solution and  mass point.
The weight can be written as $\mathrm{ \sum  F(x_{i}) F(\bar{x_{i}}) }$, where the summation is over the parton combinations and 
$\mathrm {F}$ is the CTEQ6.1 PDF set  with momentum transfer  $\mathrm{Q^{2}=m^{2}_{t}}$
 ~\cite{cteq61}. For the estimation of the PDF values the LHAPDF interface was used ~\cite{lhapdf}.


The PDF weight normalized to unit volume provides an upper bound for  the mass values of both $\mathrm{m_{t}}$ and $\mathrm{m_{W}}$.
An example of solvability weighted by the PDFs for the same single top pair event is plotted in  Figure \ref{fig:singleevent_sol} (right).
It is interesting to mention that the prefered mass point  is not the one with the lowest  $\mathrm{m_{t}}$ and $\mathrm{m_{W}}$
values as one might have guessed from the fact that PDFs favour lower mass values. 
Use of the solvability together with a matrix element weight which depends on the model has been
proposed  for top mass estimation in a single mass dimension ~\cite{matrixweighting}. This proposal has evolved to
the matrix element weighting method in Tevatron ~\cite{D0topmass}. What is proposed in this study is a general multi-dimensional search
method in mass space using only model independent PDF weights rather than a  top mass measurement method using  matrix elements.

Detector effects can change the momenta of the leptons and jets making a solvable event not-solvable.
In many cases solvability can be recovered by smearing the leptons and jets according to detector
resolution. 
For each initial event, N test events can be created by smearing the leptons, jets and missing energy components of the recorded event 
according to the detector resolutions.
For these test events, solvability can be defined  as the fraction of them for which a specific solution exist.
Solvability of the 300 test events created by the  initial single top pair event is plotted in Figure \ref{fig:singleevent_300} (left).

\begin{figure}[t!]
\includegraphics[scale=0.39]{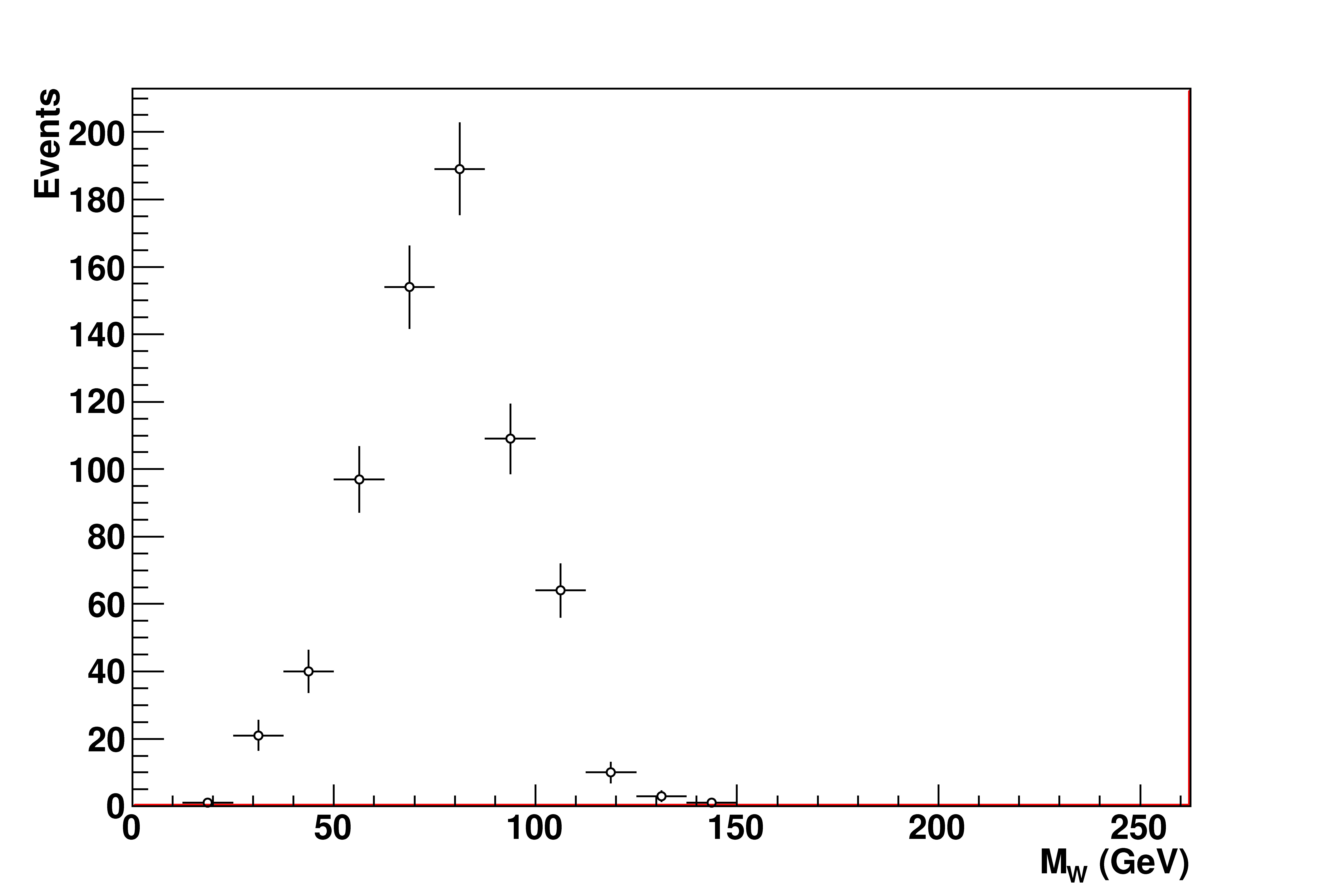}
\includegraphics[scale=0.39]{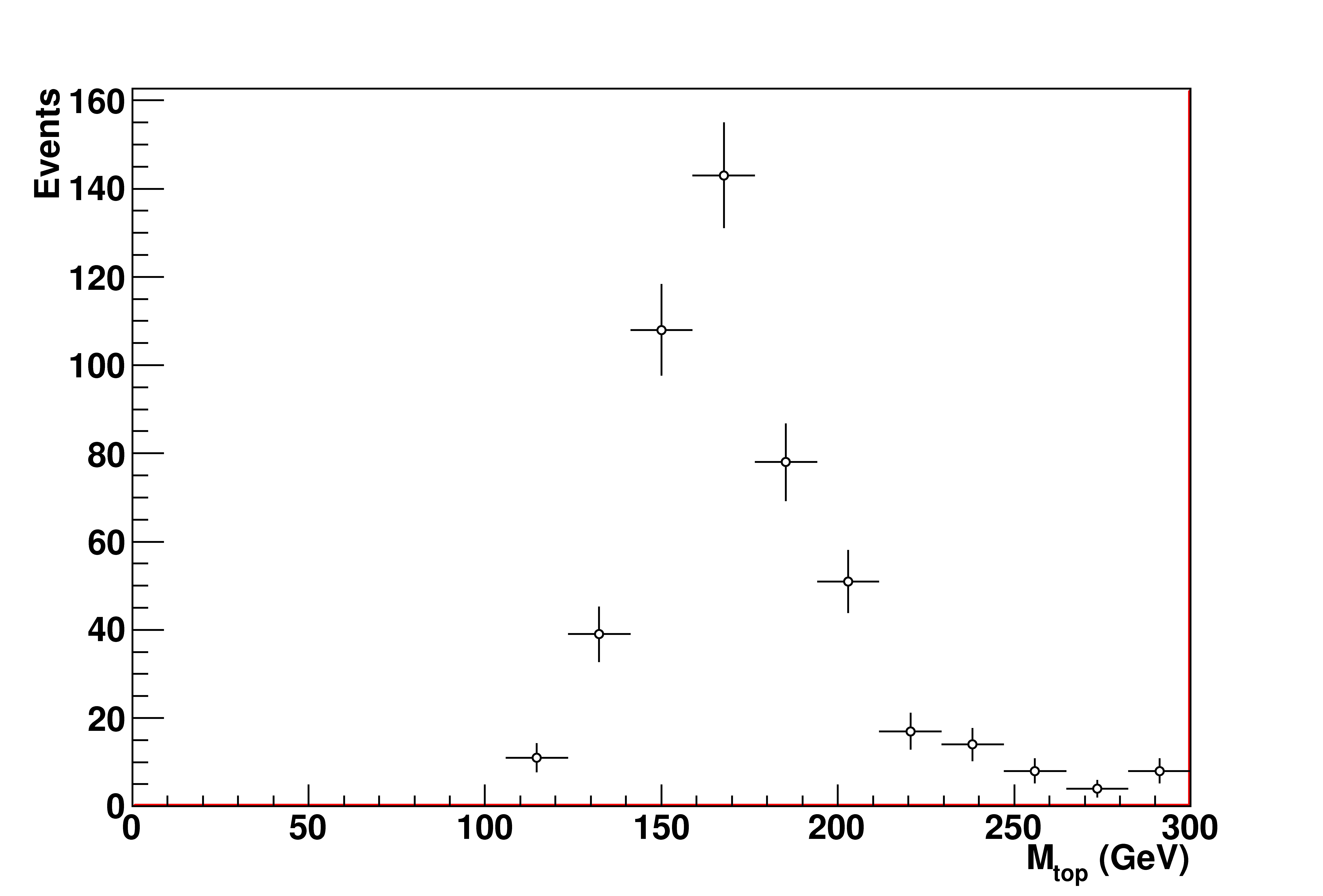}
\caption{
Reconstructed  $\mathrm{m_{W}}$ when requiring  $\mathrm{m_{t}}$ to be in the range 150-190 GeV (left)
as well as reconstructed   $\mathrm{m_{t}}$ when requiring   $\mathrm{m_{W}}$ to be in the range  70-90 GeV (right),
for a sample of signal and background events corresponding to an intergrated luminosity of  1 $\mathrm{fb^{-1}}$.
}
\label{fig:toppairsmasses}
\vspace{2em}
\end{figure}


For each test event, solvability of a solution (which is either zero or one) can be multiplied with the PDF weight calculated for the
specific solution and mass point.
The value obtained is averaged over all 300 test events and normalized to unit volume. Such a distribution can be 
constructed for all possible solutions.  An example is given in Figure \ref{fig:singleevent_300} (right) for the 
correct solution of the initial top pair event. Among all solutions, the one with the highest PDF weight is chosen.
The  final $\mathrm{m_{t}}$ and $\mathrm{m_{W}}$ estimation for this event is the mass point where the distribution of the prefered solution 
is maximized (Figure \ref{fig:singleevent_3D_300}).
The above  procedure gives a single mass  point per event.
The other option is to construct a complicated likelihood in order to exploit
all the available information. A single  mass entry is prefered so as  to reconstruct invariant mass distributions in which
robust discovery techniques can be applied. 
Is is worth emphasizing the  power of the PDF weights to choose a prefered solution. Not all solutions are as likely to originate 
from a proton-proton collision and the parton distribution functions can distinguish one of them.
This might be  applicable to other cases with combinatoric backgrounds such as reconstruction of chains with
visible particles.

\begin{figure}[b!]
\includegraphics[scale=0.38]{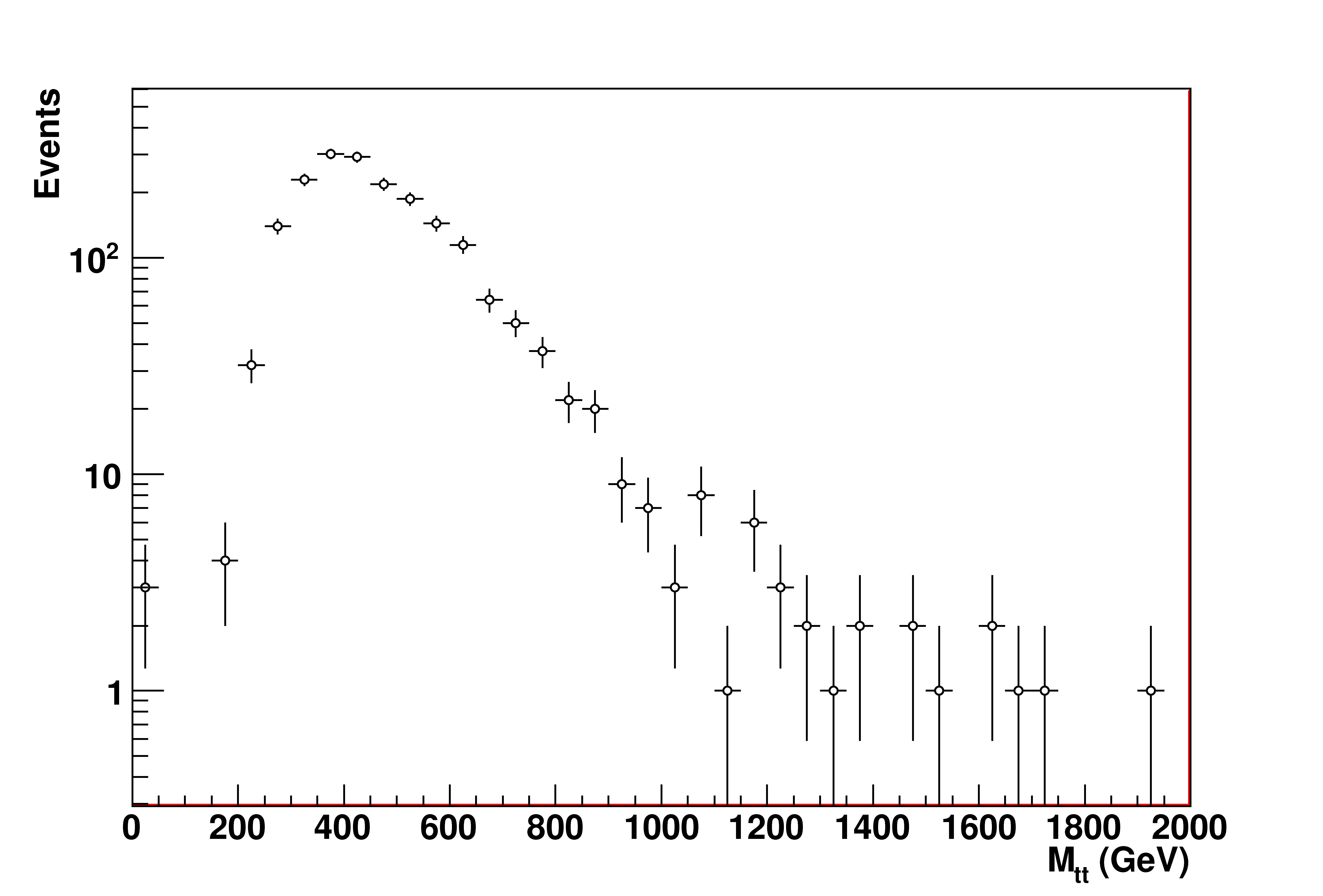}
\caption{
Reconstructed  $\mathrm{m_{t\bar{t}}}$ for a  sample of  top pair and background events corresponding to an intergrated luminosity of  1 $\mathrm{fb^{-1}}$.
}
\label{fig:smmtt} 
\vspace{2em} 
\end{figure}  

By applying the method in  signal  and background events corresponding to an  intergrated luminosity of  1 $\mathrm{fb^{-1}}$,
the  2-dimensional mass distribution  can be created (Figure \ref{fig:2D_toppairs}).
In order to produce the invariant mass distribution for the ligther particle, all events around the heavier one can be selected, according to the mass
resolution of the detector.
For W boson, all events with 150 GeV < $\mathrm{m_{t}}$ < 190 GeV in the 2-dimensional mass distribution can be selected
(Figure \ref{fig:toppairsmasses}, left) . 
In a similar way for top quark, all events with 70 GeV <$\mathrm{m_{W}}$ < 90 GeV are selected (Figure \ref{fig:toppairsmasses}, right).
The 2-dimensional mass distribution as well as the one-dimensional ones  show that the method works:
both top quark and W boson would be observed  without any a priori knowledge of their masses or of the underlying theory,
in a model independent way.
The  method described can be tested using LHC data. If it works for top pairs, it is likely to 
work for any of the searches described in the next section.

\section{APPLICATIONS}

\subsection{Search for a resonance decaying to top pairs}

An interesting application is the search for a heavy particle decaying to top pairs
$\mathrm{pp \rightarrow X'\rightarrow t\bar{t}}$ in the dilepton final state. The later has less background than the semileptonic and
hadronic top-pair final states. In addition, it allows easier boosted top reconstruction from
its final state objects as top is reconstructed from a single bjet and a lepton rather than 3 closely spaced jets.
 In order to test the method, a sample of 300 events of $\mathrm{pp \rightarrow Z'\rightarrow t\bar{t}}$ were
created, with $\mathrm{m_{Z'}}$=500 GeV and  $\mathrm{\Gamma_{Z'}}$=15 GeV. 

\begin{figure}[t!]
\includegraphics[scale=0.36]{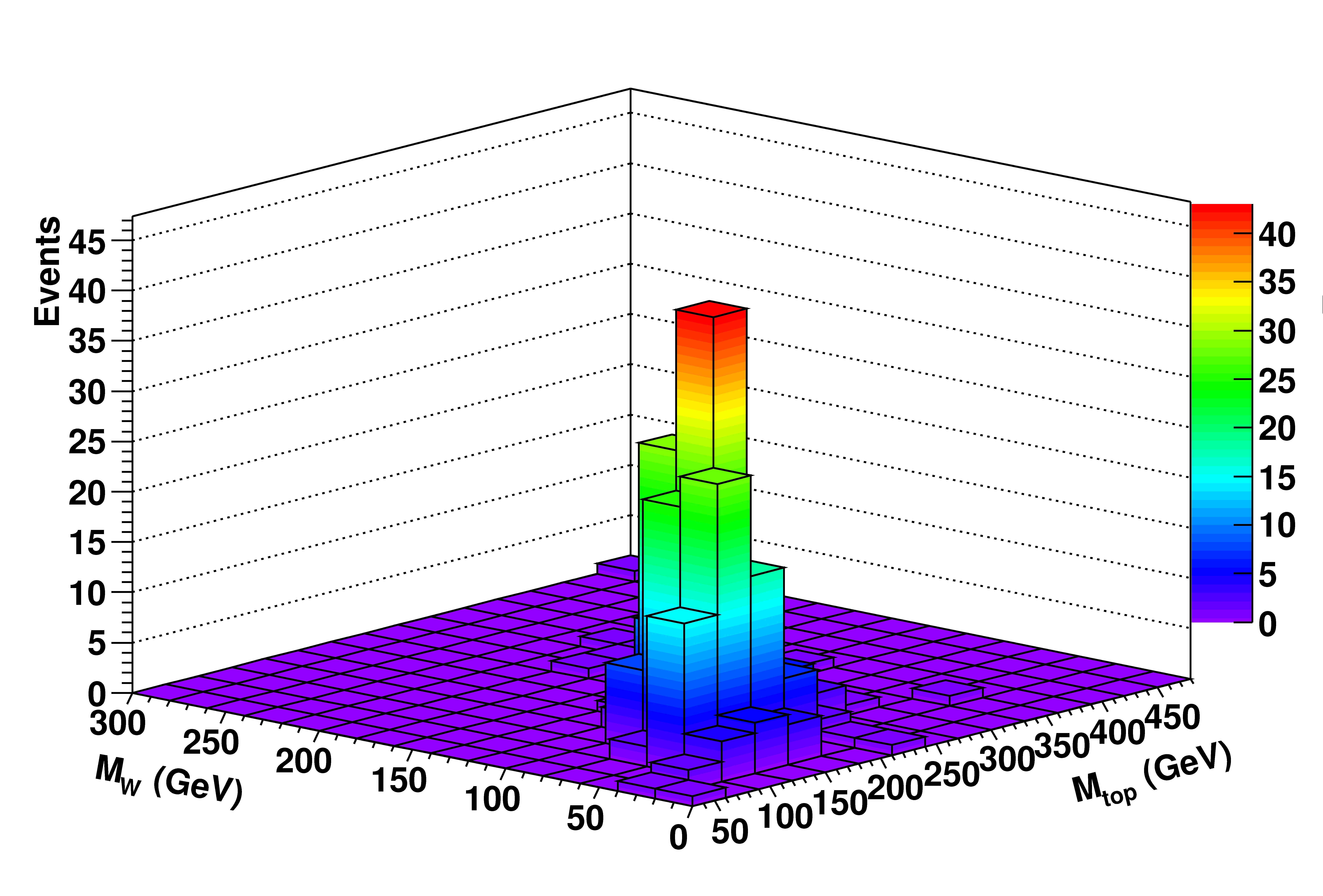}
\includegraphics[scale=0.36]{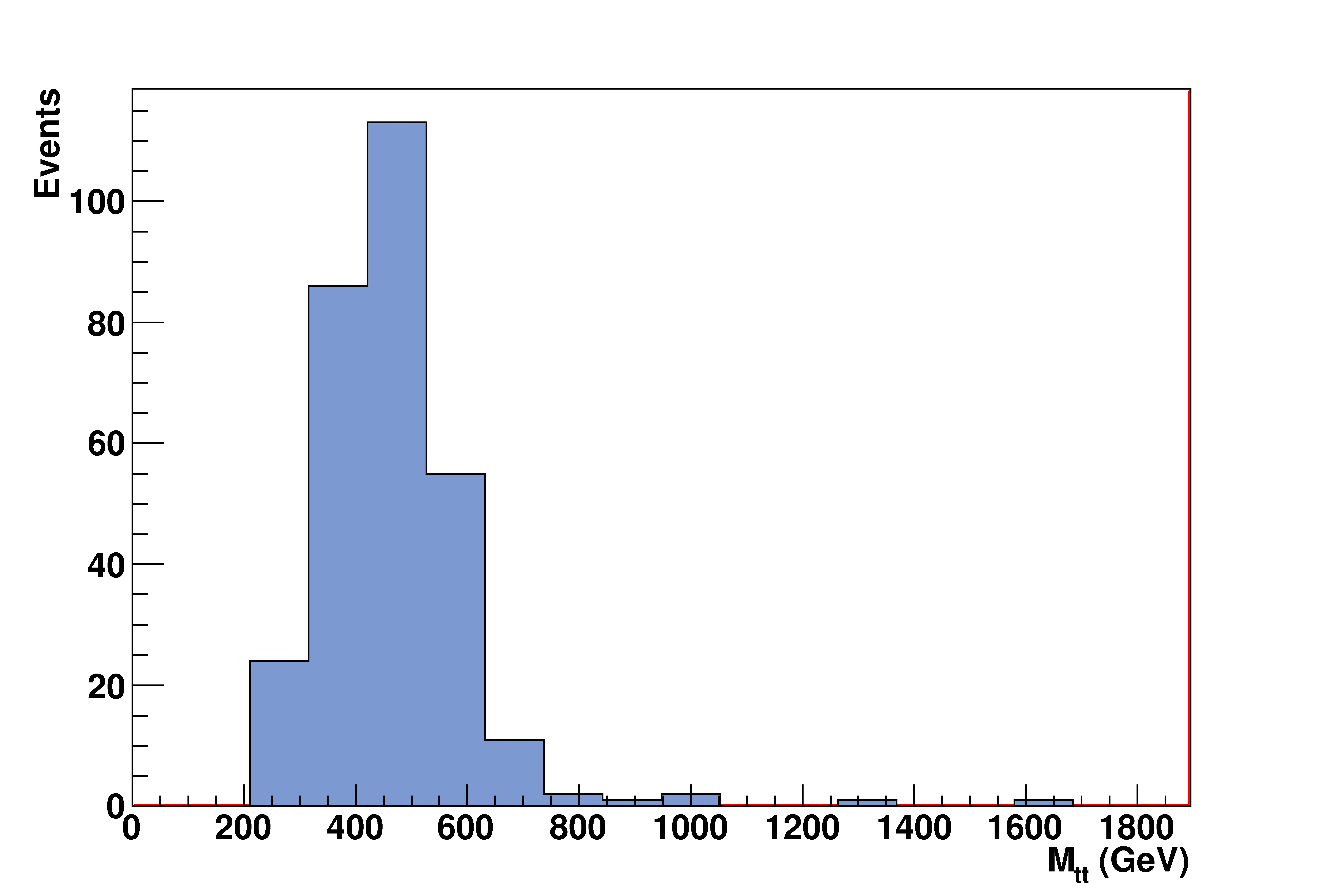}
\caption{
Reconstructed $\mathrm{m_{t}}$ and  $\mathrm{m_{W}}$  per event for a sample of 300 $\mathrm{Z'\rightarrow t\bar{t}}$  generated with
$\mathrm{m_{Z'}}$ =500 GeV (left) as well as  reconstructed  $\mathrm{m_{Z'}}$ for the same events (right).
}
\label{fig:zprime}
\vspace{2em}
\end{figure}

The average weighted solvability of the test events  is used to 
get an estimated  $\mathrm{m_{t}}$ and  $\mathrm{m_{W}}$ per event, as described in the previous section. 
The solution and mass point with the highest PDF weight is chosen.
A  full kinematic reconstruction can now take place for each of the test events:
Starting from the neutrino momenta,the four-vectors of W boson and top quark can be calculated for both branches. 
Knowledge of the above allows the  reconstruction of the  $\mathrm{t\bar{t}}$ system  and subsequently its invariant mass.
The average $\mathrm{m_{t\bar{t}}}$ of all solvable test events is used as an estimate of the mass of a possible
$\mathrm{X'\rightarrow t\bar{t}}$ decay (Figure  \ref{fig:smmtt}).
Such a resonance   would be observed
as a peak in the  $\mathrm{m_{t\bar{t}}}$ distribution, provided that its width is not very wide.
The 2-dimensional reconstructed  $\mathrm{m_{t}}$ and  $\mathrm{m_{W}}$  per event  can be seen in Figure 
 \ref{fig:zprime} (left). The reconstructed  $\mathrm{m_{Z'}}$ for the same events  is shown in Figure  \ref{fig:zprime} (right).
The invariant mass distribution is an interesting observable to look for a heavy new particle as
there are very few Standard Model events in the high mass range.

\begin{figure}[t!]
\includegraphics[scale=0.36]{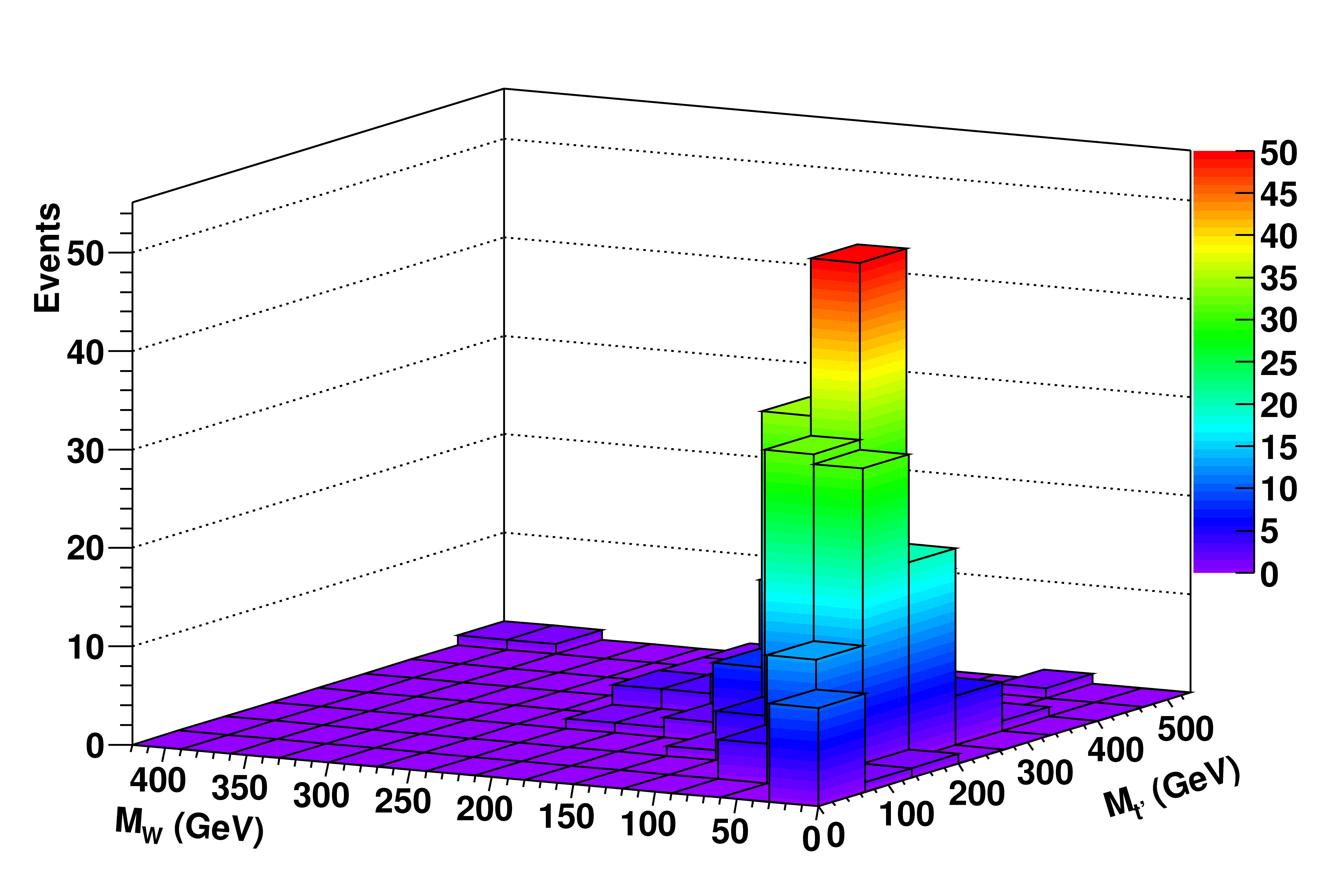}
\includegraphics[scale=0.36]{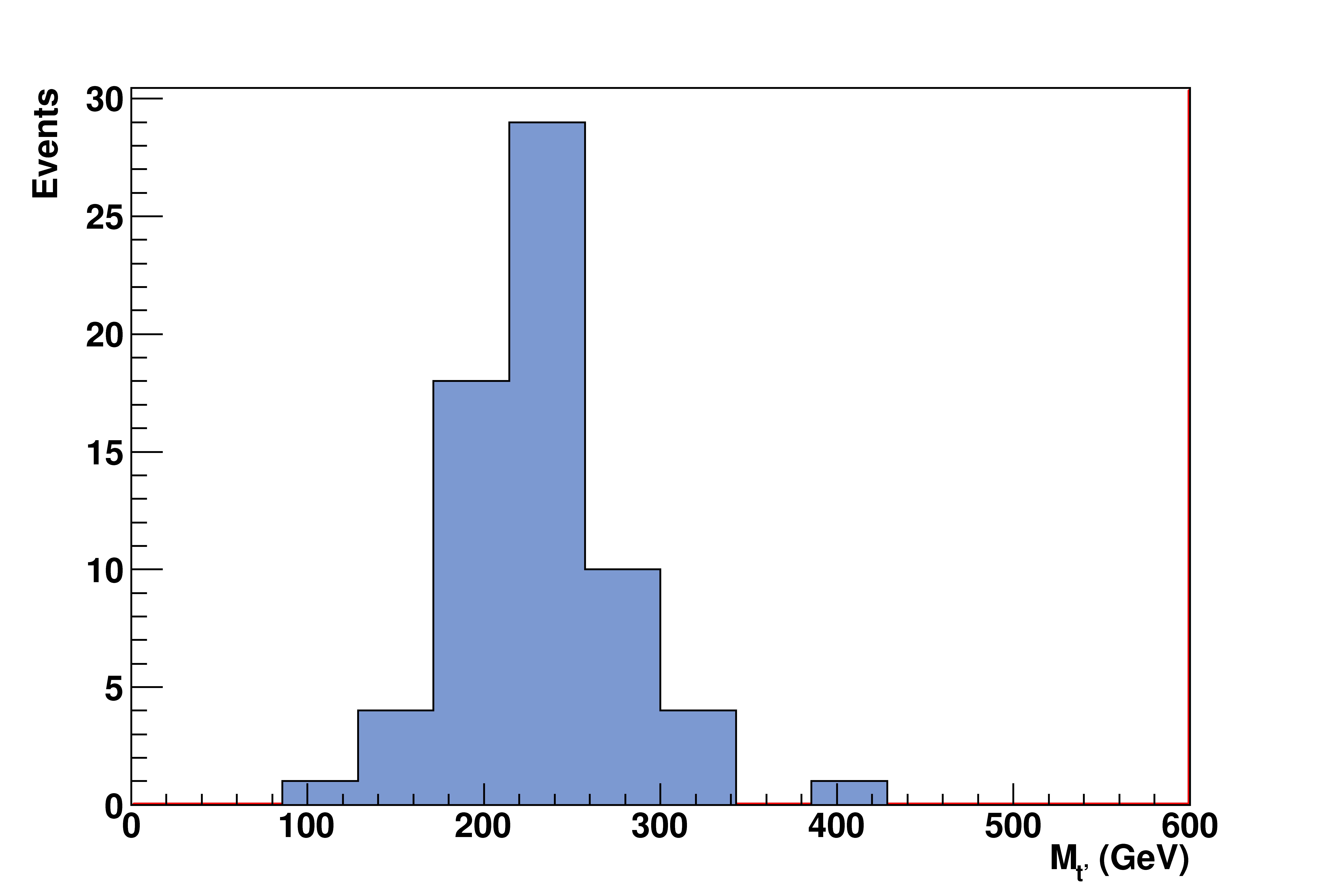}
\caption{
Reconstructed $\mathrm{m_{t'}}$ and  $\mathrm{m_{W}}$  per event for a sample of 300 
$\mathrm{t'\bar{t'}\rightarrow W^{+}b W^{-}\bar{b} \rightarrow l^{+}l^{-} b\bar{b}\nu_{l} \bar{\nu_{l}}}$ events
generated with  $\mathrm{m_{t'}}$ =250 GeV (left) as well as reconstructed  $\mathrm{m_{t'}}$ when requiring  $\mathrm{m_{W}}$ to be
in the range 70-90 GeV (right). 
           }
\label{fig:tprime}
\vspace{2em}
\end{figure}

\subsection{Search for a new generation heavy quark}

The 2-dimensional mass reconstruction descibed is model independent and would
work for any particle decaying like top quark or W boson. So one could look for anything
decaying as $\mathrm{YY\rightarrow Xb Xb \rightarrow ll bb\nu_{l} \nu_{l}}$ ($\mathrm{Y \rightarrow Xb}$,
$\mathrm{X \rightarrow l\nu_{l}}$). 
The two chains can be identical or a particle anti-particle pair  as in both cases the masses are  the same.  

In case Y is a next generation heavy quark t' the topology would be 
$\mathrm{t'\bar{t'}\rightarrow W^{+}b W^{-}\bar{b} \rightarrow l^{+}l^{-} b\bar{b}\nu_{l} \bar{\nu_{l}}}$,
($\mathrm{t' \rightarrow W b}$). Another possibility is to search for X particle instead of Y, where  X could be a charged higgs boson.
 To demonstrate the reconstruction of a $\mathrm{t'\bar{t'}}$ signal, a sample of 300 events 
were used with $\mathrm{m_{t'}}$=250 GeV. The 2-dimensional mass reconstruction can be seen in
Figure  \ref{fig:tprime} (left). The one-dimensional t' mass reconstruction can be seen in Figure  \ref{fig:tprime} (right).

\subsection{Top pair identification}

Another possible application of the 2-dimensional mass reconstruction is the identification of top pair events.
For many searches performed in final states with missing energy  top pairs are the most significant 
SM background. The $\mathrm {M_{ET}}$-like observables used to establish a discovery have their tail populated by 
top pair events. So, by identifying them using mass constraints in the 2-dimensional  $\mathrm{m_{t}}$ and
  $\mathrm{m_{W}}$ plane, we can supress significantly the most important source of a possible 
fake discovery.

\subsection{Top mass measurement}

The 2-dimensional mass reconstruction can also be used for an imporoved top mass measurement.
The simultaneous reconstruction  of $\mathrm{m_{W}}$   together with   $\mathrm{m_{t}}$
gives a handle to control the main systematic effect of the measurement: the jet 
energy scale. Calibration of this scale could be based on $\mathrm{m_{W}}$ reducing significantly
the effect on the reconstruction of  $\mathrm{m_{t}}$.

\section{Conclusions}

Mass space is the natural  space to search for new particles. Observation of mass peaks above 
background allows robust discovery using data-driven background estimation from the sidebands.
But most important, reconstruction of the unknown masses gives valuable insights to what the
new physics is. The search is model independent as the only assumption is the decay topology.
As a proof of principle, mass peaks of both top quark and W boson can be produced from 
leptonic top pair decays from simulated events as well as using LHC data. 
It is shown that the method can be used to reconstruct mass peaks from heavy resonances
decaying to top pairs as well as next generation heavy quarks decaying with the same event 
topology. Possible applications to supersymmetric mass reconstruction will be described in a future publication.
Other applications include top pair identification in order to reject them from
the tail of $\mathrm {M_{ET}}$-like observables for supersymmetric searches, as well as use of the
$\mathrm{m_{W}}$ to control the jet energy scale for an improved top mass measurement.



\begin{theacknowledgments}
I would like to thank Lars Sonnenschein for providing  the software used in  \cite{lans2}. 
Many thanks to Andrei Ostapchouk and Bob McErlath for discussions and suggestions concerning the PDF weights.
\end{theacknowledgments}

\appendix
\section{APPENDIX}


The top pair event topology has the following constraints:

\begin{tabular}{c  c  c}
  & &  \\ \\
  &  $  \mathrm{M_{ET_{x}}}     =   \mathrm{p_{\nu_{x}}+p_{\bar{\nu}_{x}}} $ &   $  \mathrm{E^{2}_{\nu} }   =   \mathrm{p^{2}_{\nu_{x}}}+  \mathrm{p^{2}_{\nu_{y}}} +  \mathrm{p^{2}_{\nu_{z}} }$  \\ \\
  &  $  \mathrm{M_{ET_{y}}}   =  \mathrm{p_{\nu_{y}}+p_{\bar{\nu}_{y}}} $    &   $  \mathrm{E^{2}_{\bar{\nu}} }  =  \mathrm{p^{2}_{\bar{\nu}_{x}}}+  \mathrm{p^{2}_{\bar{\nu}_{y}}} + \mathrm{ p^{2}_{\bar{\nu}_{z}}}$  \\ \\

\end{tabular}

\begin{eqnarray*}
\mathrm{m^{2}_{W^{+}}}   =  \mathrm{(E_{l^{+}}+E_{\nu})^{2}} - \mathrm{(p_{l^{+}_{x}}+p_{\nu_{x}})^{2}}\\
                           -\mathrm{(p_{l^{+}_{y}}+p_{\nu_{y}})^{2}} - \mathrm{(p_{l^{+}_{z}}+p_{\nu_{z}})^{2}} \\ \\ 
\mathrm{m^{2}_{W^{-}}}   =  \mathrm{(E_{l^{-}}+E_{\bar{\nu}})^{2}} - \mathrm{(p_{l^{-}_{x}}+p_{\bar{\nu}_{x}})^{2}} \\
                           - \mathrm{(p_{l^{-}_{y}}+p_{\bar{\nu}_{y}})^{2}}  - \mathrm{(p_{l^{-}_{z}}+p_{\bar{\nu}_{z}})^{2}}
\end{eqnarray*}

\begin{eqnarray*}
\mathrm{ m^{2}_{t} }   =  \mathrm{(E_{b}+E_{l^{+}}+E_{\nu})^{2}} - \mathrm{(p_{b_{x}}+p_{l^{+}_{x}}+p_{\nu_{x}})^{2}} \\
                        - \mathrm{(p_{b_{y}}+p_{l^{+}_{y}}+p_{\nu_{y}})^{2}}  - \mathrm{(p_{b_{z}}+p_{l^{+}_{z}}+p_{\nu_{z}})^{2}} \\ \\ 
\mathrm{m^{2}_{\bar{t}} }  =  \mathrm{(E_{\bar{b}}+E_{l^{-}}+E_{\bar{\nu}})^{2} }          - \mathrm{(p_{\bar{b}_{x}}+p_{l^{-}_{x}}+p_{\bar{\nu}_{x}})^{2}} \\
                         - \mathrm{(p_{\bar{b}_{y}}+p_{l^{-}_{y}}+p_{\bar{\nu}_{y}})^{2}} -\mathrm{(p_{\bar{b}_{z}}+p_{l^{-}_{z}}+p_{\bar{\nu}_{z}})^{2}}   \\ \\
\end{eqnarray*}



\bibliographystyle{aipproc}   

\bibliography{sample}

\IfFileExists{\jobname.bbl}{}
 {\typeout{}
  \typeout{******************************************}
  \typeout{** Please run "bibtex \jobname" to optain}
  \typeout{** the bibliography and then re-run LaTeX}
  \typeout{** twice to fix the references!}
  \typeout{******************************************}
  \typeout{}
 }

\end{document}